\documentclass[epj]{svjour}

\usepackage{amsmath,amssymb}
\usepackage{xcolor}
\usepackage{pdfcomment}
\usepackage{graphicx}
\usepackage{siunitx}
\usepackage{hyperref} % package for color links
\hypersetup{
  colorlinks,
  citecolor=blue,
  linkcolor=blue,
  urlcolor=blue
  }

\let\bs\boldsymbol

\DeclareMathOperator{\smol}{\Omega}
\DeclareMathOperator{\dsmol}{\delta\Omega}
\newcommand{\tlname}[1]{\ensuremath{\mathit{#1}}}
\newcommand{\Pe}{\tlname{Pe}}

\def\TODO[#1]#2#3{\pdfmarkupcomment[open=true,color=yellow,author={#1}]{#3}{#2}}
\makeatletter
\AtBeginDocument{\immediate\openout\ulp@out=\jobname.upa\relax}
\AtEndDocument{\immediate\write\@auxout{\string\ulp@afterend}}
\makeatother

\usepackage{glossaries}
\newacronym{ABP}{ABP}{active Brownian particle}
\newacronym{MCT}{MCT}{mode-coupling theory of the glass transition}
\newacronym{ABPMCT}{ABP-MCT}{mode-coupling theory for active Brownian particles}
\newacronym{ITT}{ITT}{integration-through transients}
\newacronym{MSD}{MSD}{mean-squared displacement}
\newacronym{BD}{BD}{Brownian dynamics}
\newacronym{EDBD}{ED-BD}{event-driven Brownian dynamics}
\newacronym{DFT}{DFT}{density-functional theory}
\newacronym{MHNC}{MHNC}{modified hypernetted-chain}
\newacronym{AOUP}{AOUP}{active Ornstein-Uhlenbeck particle}
\newacronym{SE}{SE}{Stokes-Einstein}
\newacronym{MIPS}{MIPS}{motility-induced phase separation}

\begin{document}

\title{Transport Coefficients in Dense Active Brownian Particle Systems: Mode-Coupling Theory and Simulation Results}
%\subtitle{}

\author{Julian Reichert\inst{1} \and Leon F. Granz\inst{1} \and Thomas Voigtmann\inst{1,2}}

\titlerunning{Transport Coefficients in Active Brownian Particles}
\authorrunning{Reichert, Granz, and Voigtmann}

\institute{%
Institut f\"ur Materialphysik im Weltraum,
  Deutsches Zentrum f\"ur Luft- und Raumfahrt (DLR), 51170 K\"oln,
  Germany
\and
Department of Physics,
  Heinrich-Heine Universit\"at D\"usseldorf,
  Universit\"atsstr.~1, 40225 D\"usseldorf, Germany
}

\date{Received: \today / Revised version: \today}

\abstract{
We discuss recent advances in developing a mode-coupling theory of the
glass transition (MCT) of two-dimensional systems of active Brownian particles
(ABP). The theory describes the structural relaxation close to the
active glass in terms of transient dynamical density correlation functions.
We summarize the equations of motion that have been derived for the
collective density-fluctuation dynamics, and those for the tagged-particle
motion. The latter allow to study the dynamics of both passive and active
tracers in both passive and active host systems. In the limit of small
wave numbers, they give rise to equations of motion describing the
mean-squared displacements (MSD) of these tracers and hence the long-time
diffusion coefficients as a transport coefficient quantifying long-range
tracer motion. We specifically discuss
the case of a single ABP tracer in a glass-forming passive host suspension;
a case that has recently been studied in experiments on colloidal Janus
particles. We employ event-driven Brownian dynamics (ED-BD) computer
simulations to test the ABP-MCT, and find good agreement between the two
for the MSD, provided that known errors in MCT already for the passive
system (i.e., an overestimation of the glassiness of the system) are accounted
for by an empirical mapping of packing fractions and host-system self
propulsion forces. The ED-BD simulation
results also compare well to experimental data, although a peculiar
non-monotonic mapping of self-propulsion velocities is required.
The ABP-MCT predicts a specific self-propulsion dependence of the Stokes-Einstein
relation between the long-time diffusion coefficient and the host-system
viscosity that matches well the results from simulation.
An application of ABP-MCT within the integration-through transients (ITT)
framework to calculate the density-renormalized effective swim velocity
of the interacting ABP agrees qualitatively with the ED-BD simulation
data at densities close to the glass transition, and quantitatively
for the full density range only after the mapping of packing fractions
employed for the passive system.
\PACS{
  {PACS-key}{describing text of that key} \and
  {PACS-key}{describing text of that key}
}
}

\maketitle

\section{Introduction}

The study of transport phenomena far from equilibrium is a current exciting
topic in statistical physics. One class of non-equilibrium systems is provided
by living matter, defined as those biological systems where on the microscopic
level, some mechanism is present to convert energy supplied by some fuel or
food into directed motion.
In microswimmer suspensions, these ``active'' or ``self-propelled'' entities
are of some $\si{\micro\meter}$ in size and are thus subject to both
thermal-equilibrium fluctuations that cause Brownian motion, and a motility
that is caused by their non-equilibrium driving forces
\cite{Ramaswamy.2010,Elgeti.2015,Ramaswamy.2017}.
This alone causes an interesting interplay of dynamical effects;
even more intriguing is this interplay in systems of interacting microswimmers,
or in systems where microswimmers interact with ordinary ``passive''
Brownian particles.

Detailed experimental studies of interacting microswimmers are possible
in colloidal suspensions of Janus particles \cite{Howse.2007,Baraban.2012,Bechinger.2016}.
These are generally
particles that have two chemically different sides, such that a specifically
designed interaction with the solvent can trigger phoretic forces
causing motility.
A specific example are colloidal particles coated with a light-absorbing
surface in a suspension where local heating causes reversible micro-scale
phase separation in the solvent \cite{Volpe.2011,Buttinoni.2012}.
This model system has been studied extensively \cite{Bechinger.2016,GomezSolano.2017}.
It has also, together with computer simulations, established one of the
most remarkable effects that appears in the moderately dense suspension
of active particles, viz.\ that of \gls{MIPS} \cite{Buttinoni.2013,Bechinger.2016}. It is generally believed
that the equilibrium direct interaction between the particles is hard-sphere
like, and hence the observation of phase separation (or cluster formation)
in a system with no apparent attractive interactions is rather striking.

More recently, experimental research has turned to very dense systems
close to dynamical arrest at a glass transition and to the interaction of
active particles with viscoelastic surroundings. We refer to a recent review
by Janssen for an excellent overview
\cite{Janssen.2019}.
Such situations might be closer to biophysical relevance, because most
bio-relevant active particles tend to move in environments that are
``crowded''. In particular, in experiment, the motion of a single active Janus particle
in a suspension of passive particles has been studied through
its \gls{MSD} \cite{Lozano.2019}. There emerges an interesting sequence
of both sub-diffusive and super-diffusive motion,
which signals a competition between dynamical arrest and
persistent active motion.

Theoretical modeling of microswimmers proceeds via various model
systems, among them that of \glspl{ABP} \cite{Romanczuk.2012,Bialke.2012,Fily.2012,Fily.2013,Berthier.2013,Ni.2013,Siebert.2017}. In this model, Brownian translational
and rotational diffusion is supplemented by a fixed self-propulsion velocity
that causes the particles to move persistently with a fixed velocity
in the direction of their (changing) orientation.
Although in the dilute limit, many of the different models to implement
active motion (such as the \gls{AOUP} model and related \cite{Berthier.2014,Levis.2015,Szamel.2015,Szamel.2016,Flenner.2016,Berthier.2018,Mandal.2016,Feng.2017,McCusker.2019}) are roughly
equivalent \cite{Cates.2013}, they differ in the treatment of the coupling of orientational
motion to self-propelled translational motion and in the treatment of Brownian to active forces.
In particular,  it is not evident whether the effective treatment
of persistent motion that is encoded in these models is justified in very dense
systems: it will be once the length scale of typical swimming motion
before a particle loses memory of its initial orientation is small
enough; yet one easily imagines that in a dense suspension, the small
interparticle distance introduces a length scale that will interfere
in subtle ways with the persistence length.

This rationale prompted us to develop a \gls{MCT} to describe the approach
to dynamical arrest in a dense \gls{ABP} system, starting from the full
orientation-resolved equations of motion. While arguably more complicated
than other approaches, this \gls{ABPMCT} proved capable of describing
states of dynamical arrest that depend on both the strength and the
persistence of self-propelled motion \cite{Liluashvili.2017}.
More recently, we have extended this theory to include also equations of
motion for the tagged-particle dynamics \cite{tagged} and,
based on that, the \gls{MSD} of tracer particles \cite{msd}.
This includes the case of tracers of
different activity than that of the host system, and allows a more
direct comparison of the theory to experiment, including the prediction
of the interplay of sub- and super-diffusive motion that is not evidently
present in orientation-averaged descriptions.

In the present contribution, we summarize these recent developments
of \gls{ABPMCT}, and we provide a direct comparison of the theory to
\gls{EDBD} computer simulations of \gls{ABP} systems. Establishing the
link between theory and simulation, we further extend to compare also
the simulations to the experimental data of Ref.~\cite{Lozano.2019},
to establish the extent to which this experimental model system can be
taken as a realization of hard-sphere-like \gls{ABP}.

A further specific point of \gls{ABPMCT} is that it is based on the
\gls{ITT} approach to the calculation of non-equilibrium transport coefficients
in driven systems. Within \gls{ITT}, one derives generalized Green-Kubo
relations that link these transport coefficients to specific,
microscopically defined, transient dynamical correlation functions. Here,
the term transient correlation function is taken to mean those dynamical
correlation functions that are obtained from averages over the equilibrium
ensemble, but where the observables are propagated using the full
non-equilibrium dynamics of the system. \Gls{MCT}-like approximations
to these correlation functions provide first-principle predictions of
the non-equilibrium transport coefficients.

In the case of \gls{ABP}, the perhaps most interesting application in
the dense system is that of the effective swim velocity: even though
each \gls{ABP} is supplied with a fixed self-propulsion velocity $v_0$,
on a coarse-grained level the average motion of the particles is slowed
down due to interactions, to a density-dependent velocity $v(\phi)\le v_0$.
This quantity is a fundamental quantity for
theories of \gls{MIPS} \cite{Cates.2015,Menzel.2016,Alaimo.2018,Solon.2018}, and in fact \gls{ITT}
is one of the few systematic approaches to calculate it from the
microscopic equations of motion.

The paper is structured as follows: in Sec.~\ref{sec:mct} we
first outline the \gls{ABPMCT}, followed by a derivation of the
\gls{ITT} expression for the swim velocity in Sec.~\ref{sec:itt},
and by a description of the simulation technique in Sec.~\ref{sec:edbd}.
Our results for the \gls{MSD}, the comparison to the experimental
data, and a discussion of the non-equilibrium long-time active diffusion
coefficients are presented in Secs.~\ref{sec:msd} through \ref{sec:se}.
In Sec.~\ref{sec:vswim} we compare the swim velocities predicted by
\gls{ITT} in combination with \gls{ABPMCT} to those obtained from
computer simulation, before concluding in Sec.~\ref{sec:conclusion}.

\section{Methods}

\subsection{Mode-Coupling Theory}\label{sec:mct}

The \gls{ABPMCT} for the description of the collective dynamics in dense
\gls{ABP} systems has been derived in Ref.~\cite{Liluashvili.2017}. For
completeness, we recall the central equations of that theory.

We consider a system of $N$ \gls{ABP} in two spatial dimensions,
with positions $\vec r_k$ and orientation angle $\varphi_k$
($k=1,\ldots N$). The overdamped active-Brownian equations of motion are then
\begin{subequations}\label{eq:abp}
\begin{align}
  d\vec r_k&=\mu\vec F_k\,dt+\sqrt{2D_t}d\vec W_k+v_0\vec n(\varphi_k)\,dt\,,\\
  d\varphi_k&=\sqrt{2D_r}dW^\varphi_k\,.
\end{align}
\end{subequations}
The interaction forces $\vec F_k$ are approximated to encode hard-sphere
interactions, i.e., no two particles are allowed to overlap, and there is
no direct interaction among them else. Importantly, they are assumed to
be spherically symmetric. The mobility
$\mu=\beta D_t$ is chosen to obey detailed balance for the equilibrium
passive dynamics (where $v_0=0$, and $\beta=1/kT$ is the inverse temperature).
Here, $D_t$ and $D_r$ are the translational
and rotational diffusion coefficients, and $d\vec W_k$ and $dW^\varphi_k$
are component-wise independent Wiener processes that drive the diffusive
motion. We fix units of length and time by
the typical particle diameter $\sigma$ and $\sigma^2/D_t$.

The active driving acts along the particle's orientation vector
$\vec n(\varphi_k)=(\cos\varphi_k,\sin\varphi_k)^T$ and is proportional to a
self-propulsion velocity $v_0$
that is, in this model, fixed per particle.
The activity of the \gls{ABP} is thus controlled by two parameters,
the dimensionless self-propulsion velocity $v_0\sigma/D_t$
and its persistence time $D_t/\sigma^2D_r$. These parameters are the relevant
dimensionless parameters entering \gls{ABPMCT}; note that for low-density
\gls{ABP}, the alternative combination of parameters into
the P\'eclet number $\Pe=v_0^2/2D_rD_t$ and
the persistence length $\ell_p=v_0/D_r$ is more natural.

Equations~\eqref{eq:abp} define realizations of a non-Gaussian Markov process
whose time-dependent probability distribution function $p(\Gamma,t)$ in
the configuration space $\Gamma=\{\vec r_k,\varphi_k\}_{k=1,\ldots N}$
is given by the Fokker-Planck (Smoluchowski) equation
$\partial_t p(\Gamma,t)=\smol(\Gamma)p(\Gamma,t)$. The adjoint
Smoluchowski operator (under the ordinary $L^2$-function scalar product)
reads
\begin{equation}\label{eq:smol}
  \smol^\dagger=\sum_{k=1}^ND_t(\vec\nabla_k+\beta\vec F_k)\cdot\vec\nabla_k
  +D_r\partial_{\varphi_k}^2+v_0\vec n(\varphi_k)\cdot\vec\nabla_k\,.
\end{equation}
In particular, it can be written as $\smol^\dagger=\smol^\dagger_\text{eq}
+\dsmol^\dagger$, to separate the detailed-balance fulfilling equilibrium
dynamics and the nonequilibrium perturbation
$\dsmol^\dagger=v_0\vec n(\varphi_k)\cdot\vec\nabla_k$.
An alternative splitting that we will encounter in deriving the theory
is into the translational and rotational parts,
$\smol^\dagger=\smol^\dagger_T+\smol^\dagger_R$ with
$\smol^\dagger_R=D_r\partial_{\varphi_k}^2$.

\Gls{ABPMCT} starts from the angle-resolved density fluctuations,
$\varrho_l(\vec q)=\sum_{k=1}^N\exp[i\vec q\cdot\vec r_k]
\exp[il\varphi_k]/\sqrt{N}$. The central quantity of the theory are
the transient dynamical density correlation functions
\begin{equation}
  \Phi_{ll'}(\vec q,t)=\langle\varrho_l^*(\vec q)\exp[\smol^\dagger t]
  \varrho_{l'}(\vec q)\rangle\,.
\end{equation}
In these correlation functions, the averaging denoted by angular brackets
is performed over the equilibrium ensemble, i.e., over the Boltzmann
distribution $p_\text{eq}(\Gamma)$ that satisfies
$\smol^\dagger_\text{eq}p_\text{eq}=0$. (Note that in our study of an
infinite system in the thermodynamic limit, $p_\text{eq}$ does not depend
on the orientations as $\smol^\dagger_R$ and $\smol^\dagger_\text{eq}-\smol^\dagger_R$ commute.)
The initial value of the correlation functions is
$\Phi_{ll'}(\vec q,t)=S_{ll}(q)\delta_{ll'}$, the matrix of equilibrium
static structure factors. Since the orientations of the particles are
uncorrelated, we have that $S_{ll}(q)=1$ for all $l\neq0$. The
entry $S_{00}(q)=S(q)$ is the ordinary (hard-sphere) static structure
factor known from liquid-state theory of the passive system.
These functions and $\Phi_{00}(q,t)$ are isotropic functions $\vec q$,
setting $q=|\vec q|$, under the assumption that the system remains
statistically homogeneous and isotropic.
The correlation functions for $l,l'\neq0$ obey simple unitary transformation
rules under a rotation of $\vec q$: in particluar the quantities
\begin{equation}
  \tilde\Phi_{ll'}(q,t)=e^{i(l-l')\theta_q}\Phi_{ll'}(\vec q,t)
\end{equation}
where $\theta_q$ is the orientation angle of the vector $\vec q$,
do not depend on that orientation.

A Mori-Zwanzig projection operator calculation allows to derive equations
of motion for the density correlation functions,
\begin{multline}\label{eq:mz}
  \partial_t\tilde{\bs\Phi}(q,t)+\tilde{\bs\omega}(q)\cdot\bs S^{-1}(q)
  \cdot\tilde{\bs\Phi}(q,t)
  \\
  +\int_0^tdt'\tilde{\bs m}(q,t-t')\cdot\left(\bs1\partial_{t'}
  +\tilde{\bs\omega}_R\right)\cdot\tilde{\bs\Phi}(q,t')=\bs0\,.
\end{multline}
Here, bold symbols refer to matrices in angular-mode indices $l$.
The matrix $\omega_{ll'}(\vec q)=-\langle\varrho_l^*(\vec q)\smol^\dagger
\varrho_{l'}(\vec q)\rangle$ is split into its translational and rotational
parts, $\bs\omega(\vec q)=\bs\omega_T(\vec q)+\bs\omega_R$ given by
\begin{subequations}
\begin{align}
  \tilde\omega_{T,ll'}(\vec q)&=D_tq^2\delta_{ll'}-\frac{iv_0q}{2}S_{ll}(q)\delta_{|l-l'|,1}\,,\\
  \tilde\omega_{R,ll'}&=D_rl^2\delta_{ll'}\,.
\end{align}
\end{subequations}

For the memory kernel, $\tilde{\bs m}(q,t)=\tilde{\bs M}(q,t)\cdot\tilde{\bs\omega}_T^{-1}(q)$, \gls{ABPMCT} proposes
\begin{subequations}\label{eq:M}
\begin{multline}
  \tilde M_{ll'}(\vec q,t)\approx\frac{n}{2}\int\frac{d\vec k}{(2\pi)^2}
  \sum_{l_3l_4l_{3'}l_{4'}}
  \tilde{\mathcal V}^\dagger_{ll_3l_4}(\vec q,\vec k\vec p)\times\\ \times
  \tilde\Phi_{l_3l_{3'}}(k,t)\tilde\Phi_{l_4l_{4'}}(p,t)
  \tilde{\mathcal V}_{l'l_{3'}l_{4'}}(\vec q,\vec k\vec p)
\end{multline}
with $\vec q=\vec k+\vec p$ and vertices $\tilde{\mathcal V}=\tilde{\mathcal V}^\text{eq}$ and $\tilde{\mathcal V}=\tilde{\mathcal V}^{\dagger,\text{eq}}
+\tilde{\mathcal V}^{\dagger,\text{neq}}$, given by
\begin{align}
\tilde{\mathcal V}^{\dagger,\text{eq}}_{l,mn}(\vec q,\vec k\vec p)
&=e^{il\theta_q}\delta_{l,m+n}\tilde{\mathcal Y}^{\dagger,\text{eq}}_{mn}(\vec q,\vec k\vec p)\,,\\
\tilde{\mathcal V}^{\dagger,\text{neq}}_{l,mn}(\vec q,\vec k,\vec p)
&=e^{il\theta_q}\delta_{|l-m-n|,1}S_{ll}(q)\delta_{|l-l'|,1}
  \tilde{\mathcal Y}^{\dagger,\text{neq}}_{l-l',mn}(\vec q,\vec k\vec p)\,.
\end{align}
Here the coupling coefficients $\tilde{\mathcal Y}^\dagger$ are determined
by the equilibrium static structure factors of the system:
\begin{equation}
  \tilde{\mathcal Y}^{\dagger,\text{eq}}_{mn}(\vec q,\vec k\vec p)
  =D_t e^{-im\theta_k}e^{-in\theta_p}\left[(\vec q\cdot\vec k)c_m(k)
  +(\vec q\cdot\vec p)c_n(p)\right]
\end{equation}
with the direct correlation function $c_l(q)=\delta_{l0}c(q)$ given by
the usual relation from liquid-state theory, $S(q)=[1-nc(q)]^{-1}$.
The nonequilibrium contribution proportional to $v_0$ reads
\begin{multline}
  \tilde{\mathcal Y}^{\dagger,\text{neq}}_{l,mn}(\vec q,\vec k\vec p)
  =\frac{iv_0}2\delta_{l,m+n}e^{-im\theta_k}e^{-in\theta_p}\times\\
  \times\Bigl[ke^{-il\theta_k}S_{l+m,l+m}(k)\tilde c_{m,l+m}(k)
  \\
  +pe^{-il\theta_p}S_{l+n,l+n}(p)\tilde c_{n,l+n}(p)\Bigr]
\end{multline}
\end{subequations}
where we have set $\tilde c_{ll'}(k)=c_{ll}(k)-c_{l'l'}(k)$.

Equations~\eqref{eq:mz} to \eqref{eq:M} form a closed set of nonlinear integral
equations that constitute the backbone of \gls{ABPMCT}.
A similar set of equations holds for the tagged-particle density correlation
function \cite{tagged},
\begin{equation}
  \phi^s_{ll'}(\vec q,t)=\langle\varrho_l^{s,*}(\vec q)
  \exp[\smol^\dagger t]\varrho_{l'}(\vec q)\rangle\,,
\end{equation}
where the $N$-particle system is extended to include one additional tracer
whose density fluctuations are $\varrho^s_l(\vec q)=\exp[i\vec q\cdot\vec r_s]
\exp[il\varphi_s]$. The Mori-Zwanzig equation then reads
\begin{multline}\label{eq:mzs}
  \partial_t\tilde{\bs\phi}^s(q,t)+\tilde{\bs\omega}^s(q)\cdot
  \tilde{\bs\phi}^s(q,t)\\
  +\int_0^tdt'\,\tilde{\bs m}^s(q,t-t')\cdot\left(\bs1\partial_{t'}
  +\tilde{\bs\omega}^s_R\right)\cdot\tilde{\bs\phi}^s(q,t')=\bs0\,.
\end{multline}
The \gls{ABPMCT} memory kernel for the tagged-particle motion is
given by
$\tilde{\bs m}^s(q,t)=\tilde{\bs M}^s(q,t)\cdot\tilde{\bs\omega}_T^{s,-1}(q)$ with
\begin{subequations}\label{eq:Ms}
\begin{multline}
  \tilde M^s_{ll'}(q,t)\approx n\int\frac{d\vec k}{(2\pi)^2}\sum_{l_3l_4}
  \tilde{\mathcal W}^s_{ll',l_3l_4}(\vec q,\vec k)\times \\ \times
  \tilde\Phi_{l_30}(k,t)\tilde\phi^s_{l_4l'}(p,t)\,.
\end{multline}
In other words, the tagged-particle dynamics can be evaluated once the
collective dynamics in terms of the $\bs\Phi(\vec q,t)$ has been
determined. The coupling coefficients for the tagged-particle memory kernel
are $\tilde{\mathcal W}^s=\tilde{\mathcal W}^{s,\text{eq}}
+\tilde{\mathcal W}^{s,\text{neq}}$ with
\begin{align}
  \tilde{\mathcal W}^{s,\text{eq}}_{ll',mn}(\vec q,\vec k)&=
  e^{i(l-l')(\theta_q-\theta_p)}\delta_{ln}\delta_{m0}\tilde{\mathcal Y}^{s,\text{eq}}(\vec q,\vec k)\,,\\
  \tilde{\mathcal W}^{s,\text{neq}}_{ll',mn}(\vec q,\vec k)&=
  e^{i(l-l')(\theta_q-\theta_p)}\delta_{|l-m-n|,1}e^{i(l-n)\theta_p}
  \tilde{\mathcal Y}^{s,\text{neq}}_{l,mn}(\vec q,\vec k)\,,
\end{align}
and
\begin{align}
  \tilde{\mathcal Y}^{s,\text{eq}}(\vec q,\vec k)
  &={D_t^s}^2(\vec q\cdot\vec k)^2(c^s(k))^2\,,\\
  \tilde{\mathcal Y}^{s,\text{neq}}_{l,mn}(\vec q,\vec k)
  &=D_t^s(\vec q\cdot\vec k)\frac{ik}2e^{-i(l-n)\theta_k}(c^s(k))^2\times\nonumber\\ &\qquad\times
  \left(v_0^s\delta_{m0}-v_0S(k)\delta_{ln}\right)\,.
\end{align}
\end{subequations}
For a detailed derivation of this result, we refer to Ref.~\cite{julianphd}.
The peculiar structure in coupling of angular modes stems from the
fact that the particles are interacting isotropically.
Equations~\eqref{eq:mzs} and \eqref{eq:Ms} then are a closed set of integral
equations for the tagged-particle correlation functions, given knowledge
of the collective dynamics.

From the limit $q\to0$ in $\phi^s_{00}(q,t)\simeq1-(q^2/4)\delta r^2(t)
+\mathcal O(q^4)$ one obtains the \gls{MSD} $\delta r^2(t)$
or the tracer particle. Performing this limit in the \gls{ABPMCT}
equations is an intricate procedure made complicated by the fact that
the frequency matrix $\bs\omega_T(\vec q)$ is tri-diagonal and has to be
inverted with care in the small-$q$ limit. After a tedious calculation
one obtains a set of integral equations to determine $\delta r^2(t)$ and
its dipole counterpart $\hat\phi^s_{\pm1,0}(t)=\lim_{q\to0}(1/q)\tilde\phi^s_{\pm1,0}(q,t)$. We get
\begin{multline}\label{eq:msd}
  \partial_t\delta r^2(t)
  =4D_t^s-2\sum_\pm(iv_0^s)\hat\phi^s_{\pm1,0}(t)\\
  -\int_0^tdt'\,\hat m_{00}^s(t-t')\delta r^2(t')\\
  +4\sum_\pm\int_0^tdt'\,\hat m_{0,\pm1}^s(t-t')
  (\partial_{t'}+D_r^s)\hat\phi^s_{\pm1,0}(t)\,,
\end{multline}
and
\begin{multline}\label{eq:msd1}
  (\partial_t+D_r^s)\hat\phi^s_{\pm1,0}(t)=\frac{iv_0^s}2\\
  -2\int_0^t\hat m^s_{\pm1,\pm1}(t-t')(\partial_{t'}+D_r^s)
  \hat\phi^s_{\pm1,0}(t')\,.
\end{multline}
The memory kernels in these equations
are given by $\hat m_{ll'}^s(t)=\lim_{q\to0}\tilde m_{ll'}(q,t)/q^{|l-l'|}$
which are well-defined for $|l-l'|\le1$. An explicit calculation of the
memory kernels is given in Refs.~\cite{msd,julianphd}. Here
we just point out that to obtain the correct $q\to0$ limit of the theory,
it is crucial that one recognizes that the $\tilde{\bs\omega}^s_T(q)$
are elements of an infinite-dimensional matrix algebra, since the
angular-mode indices $l,l'\in[-\infty,\infty]$. The inversion of this
matrix which is required to evaluate $\tilde{\bs m}^s(q,t)$
has to be performed on this infinite-dimensional algebra. Only afterwards,
an angular-momentum cutoff (as required for numerical solutions of the
equations) can safely be introduced.

Equations \eqref{eq:msd} and \eqref{eq:msd1} deserve some discussion. 
In the low-density limit, all memory kernels $\hat m_{ll'}^s(t)$ vanish,
and one is left with two coupled differential equations that can be
solved analytically. One recovers then the familiar result of the
\gls{MSD} of a free \gls{ABP},
\begin{equation}\label{eq:msdfree}
  \delta r^2(t)=4D_tt\left(1+\Pe\left(1+\frac{e^{-D_rt}-1}{D_rt}\right)\right)
\end{equation}
(dropping $s$ superscripts for simplicity).
At finite host-system density, for a passive tracer particle Eq.~\eqref{eq:msd}
decouples from Eq.~\eqref{eq:msd1} as in this case $\hat\phi^s_{\pm1,0}(t)=0$
identically. There appears then a memory kernel $\hat m_{00}^s(t)$ that
is seen to have two contributions: one that reduces to the \gls{MCT}
coupling coefficients in equilibrium, where any self-propulsion of the
host-system particles only enters through the activity-enhanced relaxation
of the collective density correlation functions. This contribution effectively
describes that the passive tracer will experience a renormalized long-time
diffusion coefficient that decreases strongly with increasing host-system
density, and is increased by the host-system activity. There is a second
contribution to $\hat m_{00}^s(t)$ that is directly proportional to
the host-particle self-propulsion velocity $v_0$, and this contribution
is crucial to obtain also a regime of super-diffusive motion in the \gls{MSD}
that a passive tracer can experience due to interactions with the persistent
swimming of the host particles \cite{msd}.

In the following, we will focus on the case of an active tracer particle
in a passive host system, to also connect to recent experiments close to
the glass transition \cite{Lozano.2019}.
In this case, all three memory kernels appearing in Eqs.~\eqref{eq:msd}
and \eqref{eq:msd1} remain relevant.

For our numerical solutions of \gls{ABPMCT} we employ an expression for
the static structure factor $S(k)$ of hard disks that was derived using
\gls{DFT} \cite{Thorneywork.2018}. Since we consider tracer particles that are,
in terms of their direct interaction (but not necessarily their self-propulsion)
identical to the host particles, we set $c^s(k)=c(k)$.
The wave-number integrals are performed on a regular grid with $M=128$
grid points up to a cutoff of $q_\text{max}\sigma=40$. To reduce numerical
effort, an angular-mode cutoff $L=1$, such that $l,l'\in[-L,L]$, was introduced.
This allows for numerically stable solutions of the \gls{ABPMCT} equations
up to self-propulsion velocities $v_0\approx8\sigma/D_t$. Details of
the time-domain integration algorithm can be found in Ref.~\cite{Liluashviliphd}.

\subsection{Integration Through Transients}\label{sec:itt}

The \gls{ABPMCT} with its focus on the transient correlation functions is
suited to evaluate non-equilibrium transport coefficients following the
\gls{ITT} approach. \Gls{ITT} was pioneered in the context of shear-driven
soft-matter glasses by Fuchs and Cates \cite{Fuchs.2002c,Fuchs.2009}. A formal
integration of the Smoluchowski equation yields for the non-equilibrium
stationary average of an observable $A$
\begin{equation}\label{eq:itt}
  \langle A\rangle_\text{neq}=\langle A\rangle_\text{eq}
  +\int_0^\infty dt'\left\langle\frac{\dsmol p_\text{eq}}{p_\text{eq}}
  e^{\smol^\dagger t'}A\right\rangle_\text{eq}\,.
\end{equation}
This \gls{ITT} formula allows to calculate the change in the ensemble
average of $A$ that is caused by the change in the probability distribution
function in response to the non-equilibrium driving term $\dsmol$.
This makes it ideally suited to address the interaction-renormalization
of transport coefficients; however, not the changes caused on a single-particle
level as for example the activity-induced extra stresses and pressure
terms that occur even if the distribution function remains flat
\cite{Pototsky.2012,Winkler.2015}.

We specificially evaluate Eq.~\eqref{eq:itt} for the effective swim
velocity. The latter is defined by
\begin{equation}\label{eq:vswim}
  v(\phi)=v_0+\frac{1}{N}\left\langle\sum_{k=1}^N\mu\vec F_k\cdot\vec n_k
  \right\rangle_\text{neq}\,,
\end{equation}
and equivalently $v^s(\phi)$ for a tracer particle. Employing the
\gls{ITT} formula for the second term, one obtains
\begin{equation}\label{eq:vswimittorig}
  v(\phi)/v_0=1-\int_0^tdt'\left\langle\sum_{jk}\vec F_j\cdot\vec n_j
  e^{\smol^\dagger t'}\vec F_k\cdot\vec n_k\right\rangle_\text{eq}\,.
\end{equation}
This equation was discussed in detail in the linear-response approximation
(where one replaces $\smol^\dagger$ with the passive-equilibrium time-evolution
operator) by Sharma and Brader \cite{Sharma.2016}.
Qualitatively, this equation describes how interactions decrease the
swim velocity: the positive integral term causes $v(\phi)/v_0\le1$.
However, this form is not yet suited well for approximations at high
densities, because such approximations would easily violate the
requirement that $v(\phi)/v_0\ge0$ for symmetry reasons. We thus perform
a further exact reformulation akin to the one employed in \gls{MCT}
when rewriting the Mori-Zwanzig equations of motion to an irreducible
form: we set
\begin{equation}
  \smol^\dagger=\smol^\dagger_\text{irr}+\sum_{kj}\vec F_k\cdot\vec n_k
  \rangle(D_tv_0/N)\langle\vec F_j\cdot\vec n_j\,,
\end{equation}
and after using a Dyson decomposition of $\exp[\smol^\dagger t]$ with
this splitting of the operator, one finds \cite{julianphd},
setting
\begin{equation}
  C(t)=\left\langle\sum_{jk}
  \vec F_j\cdot\vec n_j\exp[\smol^\dagger_\text{irr}t']
  \vec F_j\cdot\vec n_k\right\rangle\,,
\end{equation}
a further reduced \gls{ITT} expression from Eq.~\eqref{eq:vswimittorig}:
\begin{equation}\label{eq:vswimitt}
  v(\phi)=\frac{v_0}{1+\int_0^\infty dt'C(t')}\,.
\end{equation}
In this equation, \gls{ABPMCT} approximations for $C(t)$ can be safely
applied, and they proceed in analogy to those performed for the
memory kernels $\bs M(\vec q,t)$ and $\bs M^s(\vec q,t)$ that govern the
density-correlation functions, by projecting the fluctuating forces
onto density-pair modes. One gets
\begin{multline}\label{eq:vswimmct}
  C(t)\approx\frac{n}{8\pi}\int dk\,\tilde{\mathcal V}^{\text{swim}}(k)
  \sum_{l'l''=\pm1}\Bigl(\tilde\Phi_{00}(p,t)\tilde\Phi_{l'l''}(p,t)\\
  +\tilde\Phi_{l'0}(p,t)\tilde\Phi_{0l''}(p,t)\Bigr)
\end{multline}
where the vertex $\tilde{\mathcal V}^\text{swim}(k)$ is given by
the equilibrium direct correlation functions \cite{julianphd}.
The \gls{ABPMCT} with this approximation thus allows to calculate the
effective swim velocities of interacting \gls{ABP}, and (by a suitable
extension of Eq.~\eqref{eq:vswimmct}) of an active tracer in an active
or passive bath.

\subsection{Brownian Dynamics Simulations}\label{sec:edbd}

The results of the theory are checked against \gls{EDBD} simulation
results. This method, described in detail for the passive Brownian
system in Ref.~\cite{Scala.2007}, and first employed for \gls{ABP}
by Ni et~al.\ \cite{Ni.2013}, is essentially a rejection-free Monte Carlo
method to generate valid configurations of hard-sphere (hard-disk) systems.
The method consists of segments of the length of a ``Brownian time step''
$\tau_B$, within each of which an event-driven molecular dynamics simulation
is performed to ensure no-overlap conditions among the particles. For this,
random velocities are generated from random Gaussian trial displacements
such that in the case of a free particle, the correct diffusive motion
is generated. To implement self-propulsion, the trial displacements are
drawn with an appropriate drift, again such as to ensure that the known
analytical results for the free \gls{ABP} are reproduced.

We employed \gls{EDBD} simulations of $N=1000$ slightly size-polydisperse
hard disks. Within the parameter ranges and time scales that we study,
no signs of crystallization or motility-induced phase separation were observed.
After equilibration runs of suitable length, the simulation gives access
to the stationary-averaged correlation functions that are the counter-parts
of the transient correlation functions obtained from \gls{ABPMCT},
and likewise the \gls{MSD}. Averages over up to $200$ independent starting
configurations were employed to improve statistics.

For the purpose of the present discussion, we ignore the difference between
stationary and transient \gls{MSD}; the good agreement between theory and
simulation (see below) justifies the assumption that for the parameter range
that we study here, the two quantities do not differ qualitatively.
A similar conclusion was drawn for the case of transient and stationary
averages in \gls{MCT} for sheared passive suspensions \cite{Krueger.2010}.
There are systematic differences that can be observed in the off-diagonal
elements of the correlation function matrices for times shorter than the
rotational persistence time $1/D_r$, as discussed in
Ref.~\cite{tagged}.

From the simluations, we also extract the average swim velocities.
While the \gls{ITT} expressions mentioned above are justified under
the assumption of a smooth pair potential, taking the hard-sphere limit
only in the final expression containing the static structure functions,
the appearance of the direct interaction forces in Eq.~\eqref{eq:vswim}
is problematic for the \gls{EDBD} scheme. Instead, we obtain the
swim velocity directly from the Monte Carlo displacements:
\begin{equation}
  v(\phi)/v_0=\frac1{\mathcal N}\frac1N\sum_{i=1}^N\sum_{t\in n\tau_B}
  \frac{\Delta\vec r_i(t)}{\tau_B}\cdot\vec n_i(t)\,,
\end{equation}
where $n$ is an integer number corresponding to an average over a few
Brownian time steps and $\mathcal N$ the corresponding normalization
term $\mathcal N=n v_0$. Here, $\Delta\vec r_i(t)$ is the actual
displacement that the \gls{EDBD} algorithm assigns to particle $i$ in
a single Brownian time step.
In the case of a free particle, $\Delta\vec r_i(t)$ contains a passive
Brownian contribution that is uncorrelated with the particle orientation,
and a term $\propto v_0\tau_B\vec n_i(t)$, so that in the non-interacting
limit $v(\phi)=v_0$ is guaranteed.

\section{Results}

\subsection{Mean-Squared Displacements}\label{sec:msd}

A particularly interesting case of the dynamics is that of a single
\gls{ABP} in a dense host system of passive particles. This system has
been studied experimentally recently \cite{Lozano.2019}.

We have recently discussed the features of the \gls{MSD} of active and passive
tracers in active and passive host systems in a comparison between
\gls{ABPMCT} and \gls{EDBD} simulations \cite{msd}.
Overall, it was found, the theory describes the simulation results
qualitatively correctly. After an empirical rescaling of the density
and, in the case of an active host system, the self-propulsion
velocity by a global factor of $\mathcal O(1)$, the agreement is also
quantitative
for not too large $v_0$. In essence, the theory predicts the slowing
down of the dynamics due to approaching glassy arrest at high density,
and a rescaling of the density accounts for a numerical error in the
predicted value of the glass-transition point. Self-propulsion is seen
to fluidize the system, so that the transition to the ``active glass''
gets delayed to higher densities \cite{Liluashvili.2017}. \Gls{ABPMCT}
somewhat underestimates the effectiveness of self propulsion as compared
to the simulations, but otherwise describes it qualitatively.

For the free particle, the \gls{MSD} displays two cross-overs at length
scales associated to the persistence of self-propulsion. From
the analytical result, Eq.~\eqref{eq:msd}, one readily infers these
length scales and the associated time scales,
\begin{subequations}
\begin{align}
  \tau_v&=\frac{4D_t^s}{{v_0^s}^2}\,, &
  \ell_v&=\frac{2D_t^s}{v_0^s}\,, \\
  \tau_p&=\tau_v(1+\Pe^s)=\tau_v+\frac{2}{D_r^s}\,, &
  \ell_p&=\ell_v+\frac{v_0^s}{D_r^s}\,.
\end{align}
\end{subequations}
For $t\simeq\tau_v$,
the short-time passive Brownian diffusion (coefficient $D_t^s$)
crosses over to a super-diffusive
regime that is indicative of persistent self-propelled motion.
For $t\simeq\tau_p$, this persistence is lost, and
the free \gls{ABP} crosses over to enhanced diffusion
with coefficient $D_t^\text{eff}=D_t^s(1+\Pe^s)$.
Since at low densities these are the only length- and time-scales relevant for
the
problem, the P\'eclet number of the tracer, $\Pe^s$ is the only
dimensionless number to quantify activity in the long-time limit.

\begin{figure}
\includegraphics[width=\linewidth]{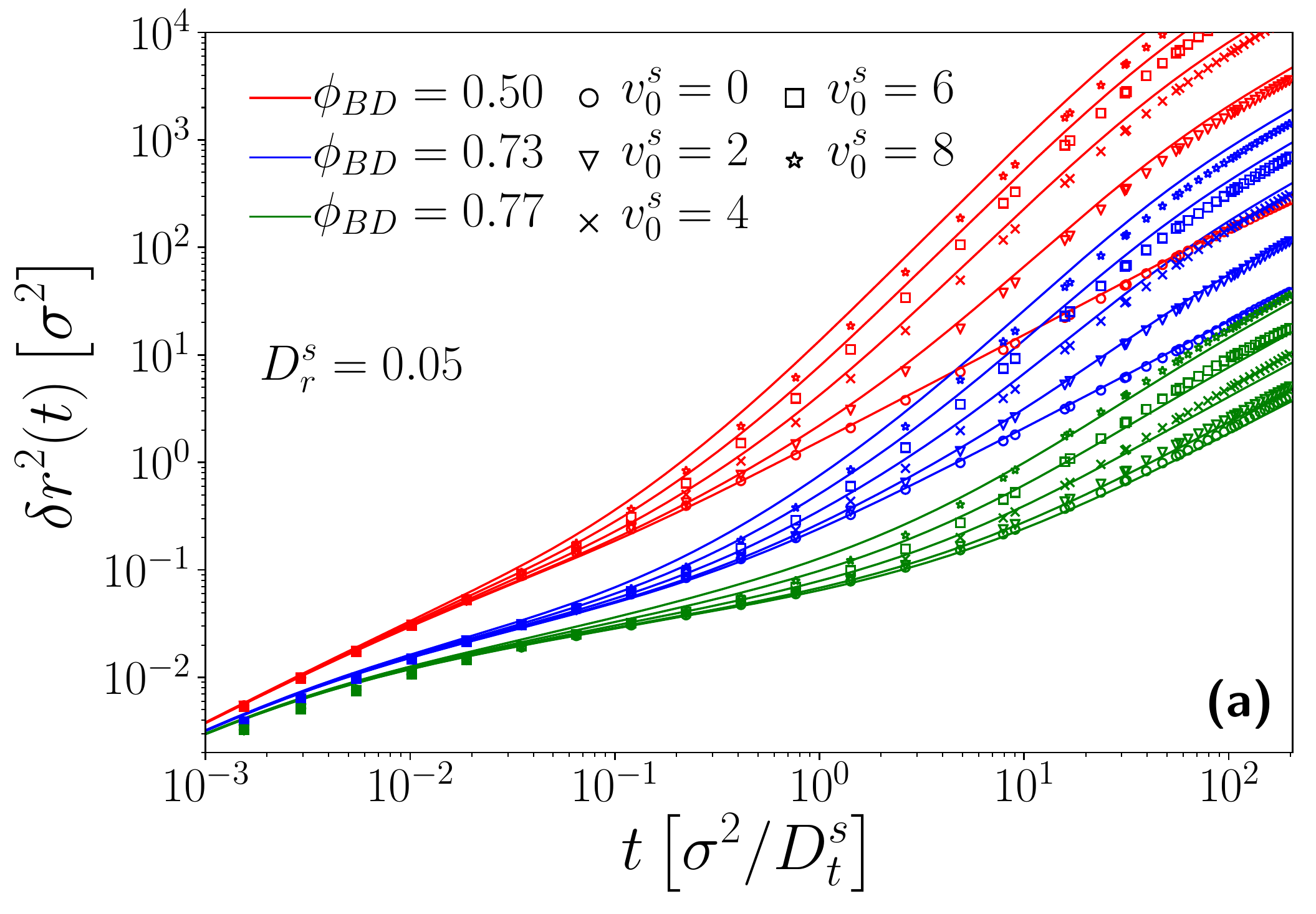}
\caption{\label{fig:434}
  Mean-squared displacement of an active tracer particle embedded in a
  host system of passive hard disks at packing fraction $\phi$ as
  labeled, for tracer-self-propulsion speeds $v_0^s$ as labeled, and
  for fixed reorientational diffusion coefficient $D_r^s=0.05$
  and with translational diffusion coefficient $D_t^s=D_t$ equal to that
  of the host system.
  Symbols are results from Brownian dynamics simulations, lines are
  results from MCT after an adjustment of the packing fraction in order
  to match the dynamics of the fully passive system.
}
\end{figure}

At high densities, the dynamics of an active particle is characterized
by a competition of time- and length-scales: particle
interactions set a length scale of nearest-neighbor cages, $\ell_c$.
For $\delta r^2(t)\simeq 4\ell_c^2$, the interactions of the tracer with
the host system cause sub-diffusive motion that ultimately leads to
dynamical arrest at the glass-transition density. Close to and on the
liquid side of the glass transition, an increasing time scale governed
by the \gls{MCT} memory kernels sets the time scale $\tau_\alpha$
on which the sub-diffusive regime ends. \Gls{ABPMCT} predicts that
$\tau_\alpha$ diverges as the glass transition is approached.
The cage length scale (of order of $10\%$ of a particle diameter)
interferes with the length scales derived from the free-\gls{ABP} motion.
Thus, there emerges a sequence of sub- and superdiffusive regimes in the
\gls{MSD} that depends on the relative magnitudes of the associated
time scales, $\tau_c=\ell_c^2/D_t^s$, $\tau_v$, and $\tau_p$, as well as
the strongly density-dependent $\tau_\alpha$.

Figure~\ref{fig:434} displays as an exemplary case the \gls{MSD}
for an active tracer various self-propulsion velocities $v_0^s\le 8\,D_t/\sigma$
in a passive bath at packing fractions approaching the glass
transition. To emphasize the effect of active motion, we have chosen
a relatively large persistence time, letting $D_r^s=0.05\,D_t/\sigma^2$.
With these parameters, we obtain for the case $v_0^s=8\,D_t/\sigma$
the relevant time scales as
$\tau_c\approx8\times10^{-3}\,\sigma^2/D_t$, $\tau_v=1/16\,\sigma^2/D_t$,
and $\tau_p=(40+1/16)\,\sigma^2/D_t$, so that $\tau_c<\tau_v\ll\tau_p$.

Correspondingly, the motion in the moderately dense host system ($\phi=0.50$
in Fig.~\ref{fig:434}) displays essentially a cross-over from short-time 
diffusion to super-diffusive persistent motion, and at $t\approx\tau_p$
a further cross-over to enhanced diffusive motion. At these densities,
the cageing influence from the host system is still too weak to be noted
dramatically, although a slight sublinear growth in the \gls{MSD} around
$t\approx\tau_c$ can be discerned.

As the host-system density is increased, $\tau_\alpha$ increases strongly,
and the sub-diffusive regime in the \gls{MSD} expands over a wider time
window. The closer one approaches the glass transition, the more the
cross-over to persistent super-diffusive motion is suppressed, so that
for the highest density shown in Fig.~\ref{fig:434} ($\phi=0.77$),
super-diffusive motion is no longer evident and the tracer activity
essentially serves to provide an enhanced long-time diffusion coefficient
as compared to the passive tracer. Note that the enhancement no longer
scales with $\Pe^s$: for the parameters used in Fig.~\ref{fig:434},
the free \gls{ABP} would show an enhacement of a factor $\Pe^s=640$
for $v_0^s=8\,D_t/\sigma$; at $\phi=0.77$ the corresponding enhancement
is only about a factor $10$.

\begin{figure}
\includegraphics[width=\linewidth]{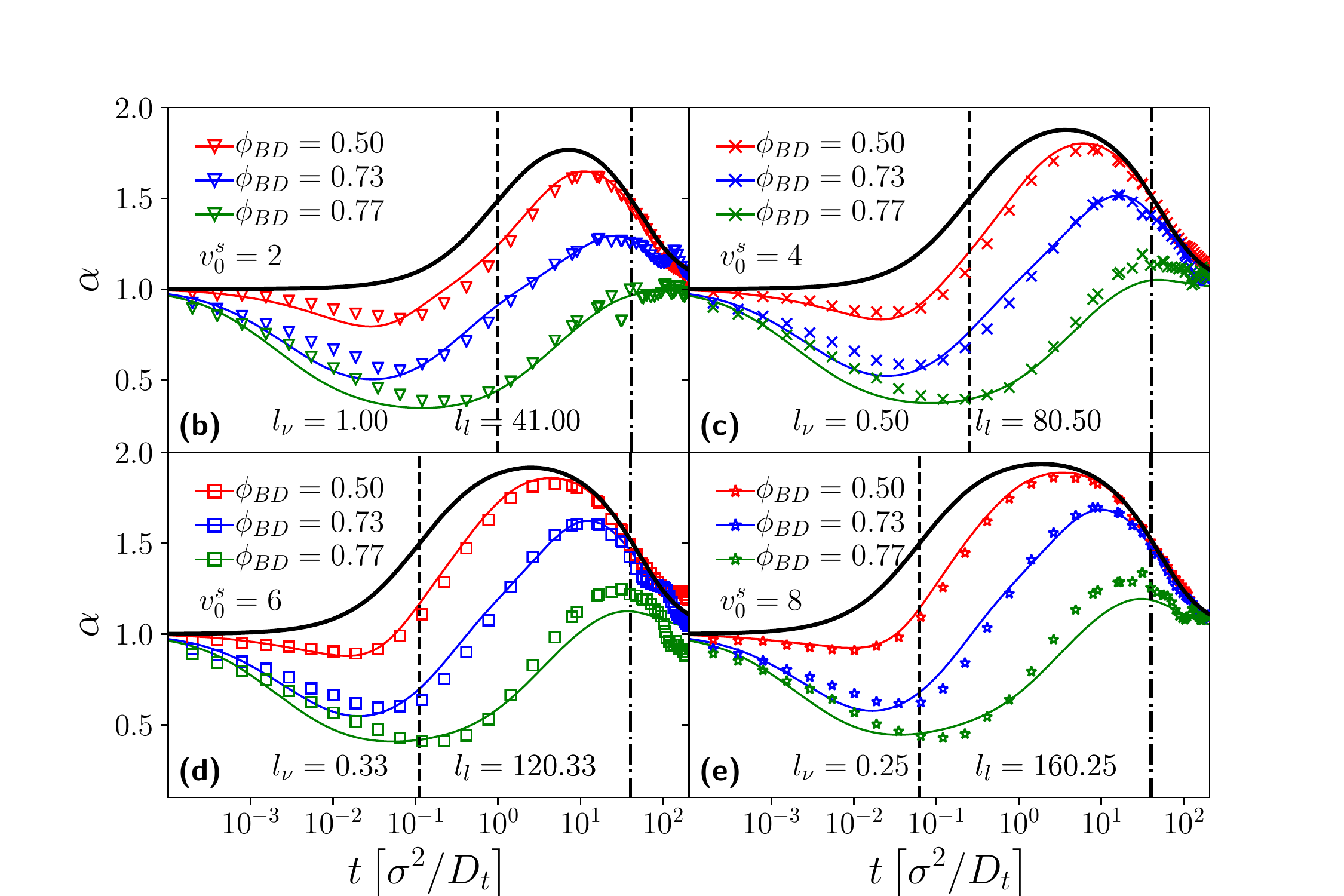}
\caption{\label{fig:434b}
  Effective exponents for the mean-squared displacements shown in
  Fig.~\protect\ref{fig:434}, obtained from
  $\alpha(t)=d\log\delta r^2(t)/d\log t$. Symbols are simulation data,
  lines results from MCT. Solid black lines are the analytical result for the
  free ABP.
  Vertical dashed and dash-dotted lines correspond to the free-ABP time
  scales $\tau_v$ and $\tau_p$, respectively.
}
\end{figure}

The sub- and super-diffusive regimes in the \gls{MSD} and their evolution
with time scales are more clearly seen in the temporal evolution of the
effective power-law exponent of the \gls{MSD}, obtained from
$\alpha(t)=d\log\delta r^2(t)/d\log t$; these are shown in
Fig.~\ref{fig:434b}. One clearly sees a decrease of $\alpha(t)$ to values
below unity around $t\approx\tau_c$, reflecting in-cage sub-diffusive
motion. The values of $\alpha(t)$ start increasing towards value above
unity at $t\approx\tau_v$, and they decrease again towards unity at
$t\approx\tau_p$. In order to see truly ``ballistic'' motion, i.e.,
$\alpha(t)\approx2$, even in the free ABP case one would have to separate
$\tau_v$ and $\tau_p$ even further, for example by further decreasing
$D_r^s$.

It should be noted that the appearance of superdiffusion in the (stationary)
MSD is a clear sign of non-equilibrium dynamics in a system whose
governing equations of motion are Markovian-stochastic. \Gls{ABPMCT}
comes at the price of being numerically demanding, but in turn it enables to
describe such dynamics correctly; more common and simpler approaches
to the glassy dynamics of \gls{ABP} often proceed by integrating out
the orientational degrees of freedom into some effective Smoluchowski
operator. It is not evident that with such approximation made from the
outset, superdiffusive \gls{MSD} can be correctly described.

\subsection{Comparison to Experiment}\label{sec:exp}

\begin{figure}
\includegraphics[width=\linewidth]{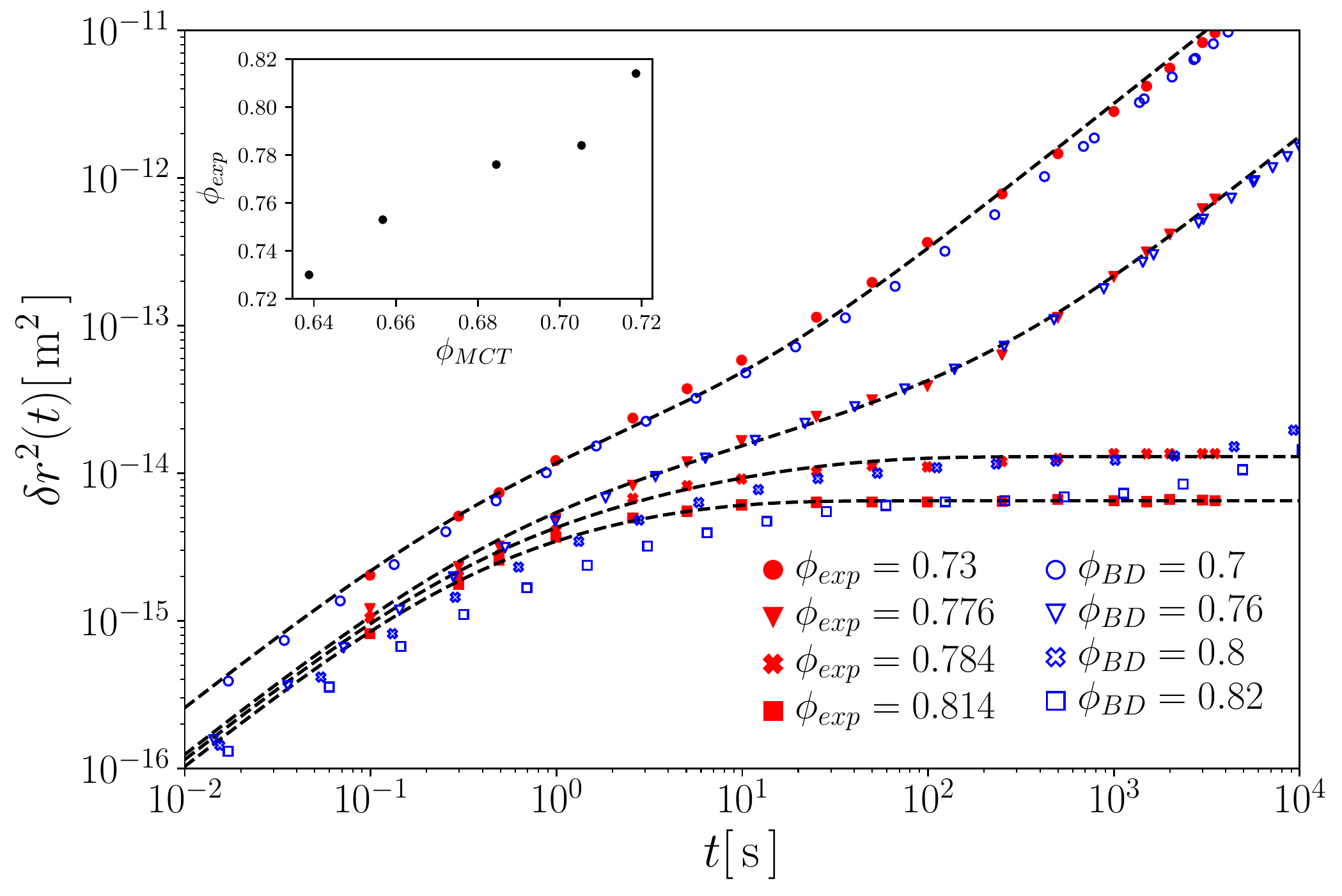}
\caption{\label{fig:522}
  Mean-squared displacement of a passive tracer in a passive host suspension,
  as obtained from a quasi-two-dimensional experiment \protect\cite{Lozano.2019}
  (filled symbols), and from our Brownian dynamics simulations
  (open symbols). Dashed lines indicate our MCT results with the
  packing fraction adjusted to best describe the simulation data.
  We find good agreement also between simulation and experiment after
  a mapping of experimental values for the packing fraction $\phi_\text{exp}$
  to that used in simulation, $\phi_\text{BD}$, that is displayed in the
  inset.
}
\end{figure}

In order to compare the recent experimental data of Lozano, Gomez-Solano, and Bechinger
\cite{Lozano.2019} with our theory and simulations, we first perform a
fit of the fully passive experimental system. This allows to establish
a precise enough mapping of packing fractions given in the experiment,
and that used in simulation. Differences are expected due to slightly
different interaction potentials (experimental particles might not interact
as idealized hard spheres), and different size polydispersity (the experimental
system uses a binary mixture).

Figure~\ref{fig:522} establishes the level of agremeent that can be achieved
between experiment, simulation, and theory for the passive dynamics.
Here, simulation and theory have been shifted by overall factors in
time and length;
With these adjustments, all three sets of data agree well with each other.
There is a deviation notable in the simulation results
around $t=1$ whose precise cause we do not know. Also, at times $t\gtrsim500$,
the \gls{BD} data at the highest packing fractions deviate from the
idealized arrest curves, most probably due to equilibration issues in the
simulation.

The rescaling in time scales serves to account for the notable effect of
hydrodynamic
interactions present in the experimental system. These cause a slowing down
already of the short-time passive diffusion that is absent in the simulation
and theory. From comparing experimental data at different packing fractions,
we estimate that with increasing $\phi$, the experimental $D_t^s(\phi)$
decreases by about a factor 5, in agreement with what is expected from
three-dimensional hard-sphere suspensions \cite{Megen.1998}.

The experimental system uses host particles of mean diameter
$\sigma\approx\SI{5}{\micro\meter}$.
From the good agreement between \gls{MCT} and \gls{EDBD} simulations
for the passive hard-disk system one expects $\ell_c\approx0.08\sigma\approx\SI{0.4}{\micro\meter}$ to hold for a system that reasonably well approximated
hard-disk behavior, for a tracer of roughly equal size to the host-suspension
particles. In the experiment, few Janus particles were added as active
tracers with a diameter $\sigma\approx\SI{6.3}{\micro\meter}$ corresponding
to that of the larger particles of the host-suspension mixture. We do not
expect the cage-localization length of these particles to be significantly
smaller than the above estimate.
There remains thus a puzzling effect in the comparison shown in
Fig.~\ref{fig:522}: the experimental data show in-cage localization around
$\delta r^2\lesssim\SI{1e-14}{\micro\meter^2}$, which corresponds to
a localization length $\ell_c^\text{exp}\approx\SI{0.05}{\micro\meter}$,
about a factor of $7$ smaller than what is estimated from theory and
simulation with the nominal hard-sphere sizes of the particles.
The reason for this discrepancy remains unclear; we proceed by adjusting
in both simulation and theory an effective diameter $\sigma_\text{eff}=\sigma/7$
of the particles that
accounts for this difference. With this adjustment, all passive experimental
data are described well.
The short-time diffusion coefficient then is read off
from Fig.~\ref{fig:522} as
$D_t\approx\SI{0.008}{\micro\meter^2\per\second}$ for the lowest density
shown in experiment.

\begin{figure}
\includegraphics[width=\linewidth]{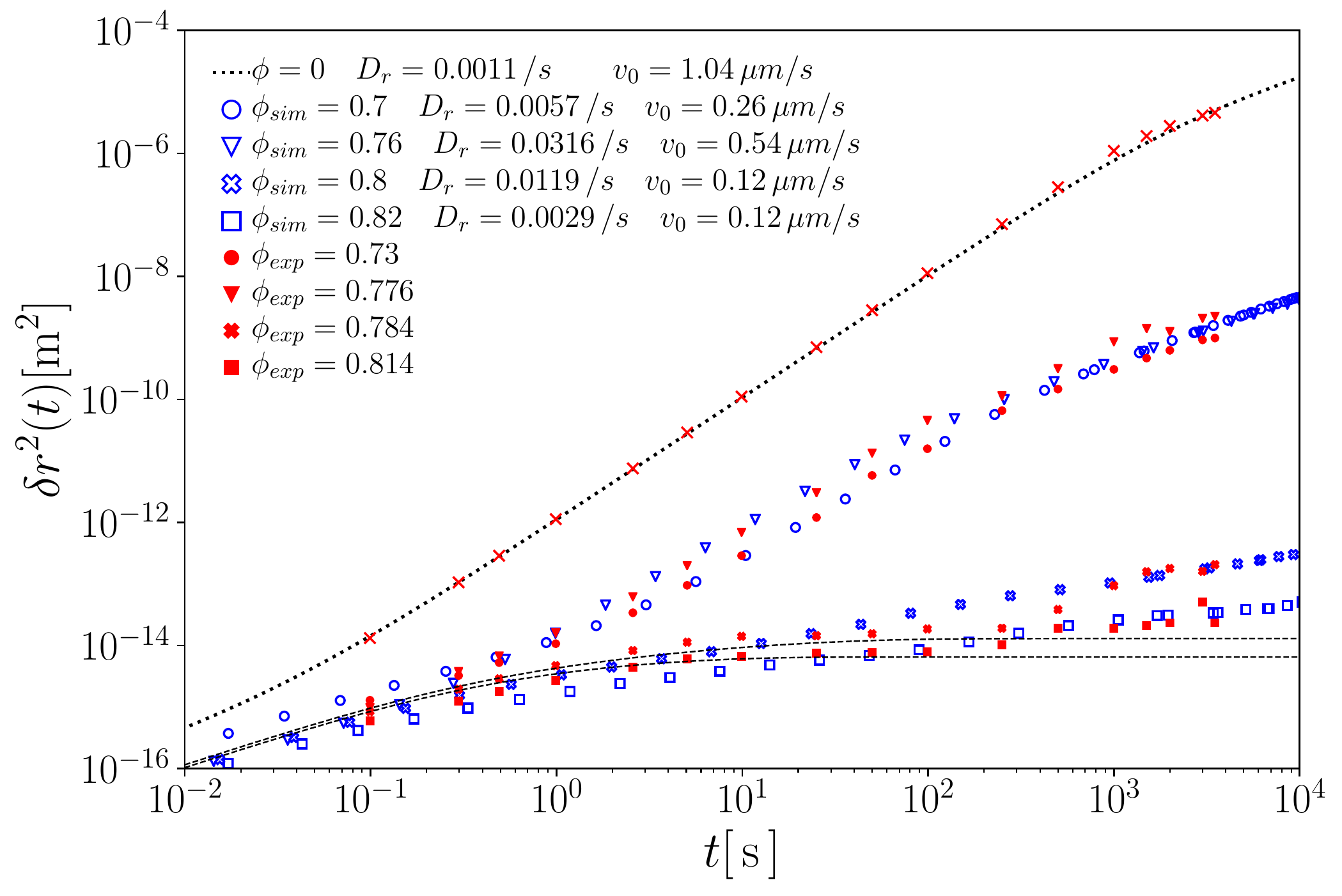}
\caption{\label{fig:531}
  Mean-squared displacements of an active tracer particle in the
  passive host system at various packing fractions $\phi$ as indicated.
  Filled symbols are taken from Ref.~\protect\cite{Lozano.2019},
  they represent the dynamics of a laser-driven Janus particle
  at fixed laser intensity.
  Open symbols are Brownian-dynamics simulation results for a hard-disk
  ABP in a system of passive hard disks, with rotational diffusion coefficient
  $D_r^s$ as determined from experiment, and self-propulsion velocity
  $v_0^s$ adjusted to
  best fit the experimental data.
  A black dotted line indicates the free-particle solution.
}
\end{figure}

For the free Janus particle, Lozano et~al.\ \cite{Lozano.2019} report
a self-propulsion velocity of $v_0^s\approx\SI{1.0}{\micro\meter\per\second}$.
Indeed, a fit of the free-\gls{ABP} \gls{MSD}, Eq.~\eqref{eq:msd}, to
the corresponding experimental data yields good agreement with
$v_0^s\sigma_\text{eff}/D_t^s=109$; this fit is shown in Fig.~\ref{fig:531}
as a black dotted line.

Our numerical algorithms to solve the \gls{ABPMCT} equations of motion
are unfortunately unstable at such large self-propulsion velocities.
For this reason, we restrict the further comparison to the experimental
data to that with \gls{EDBD} simulation results.

A remarkable effect reported in experiment \cite{Lozano.2019} is that the
reorientational dynamics of the tracer particle is strongly enhanced as
the host-system density approaches the glass transition, and then again
is suppressed in the glass. This density dependence of $D_r^s$ was
attributed to hydrodynamic coupling of the active tracer to the host
system via the solvent, and therefore it is by
definition absent in the \gls{ABP} model system. We model this
effect ``by hand'', adjusting in the simulations $D_r^s$ to the values
reported in experiment. 

After this adjustment, the long-time diffusive regime in the experimental
\gls{MSD} can be described from our simulations only after also adjusting
the self-propulsion velocity $v_0^s$ as a function of density.
The quality of agreement between experiment and simulation is then
very good, as displayed in Figure~\ref{fig:531}.
However, the values required for $v_0^s$ in the simulation show a
curious non-monotonic behavior with increasing packing fraction.

The adjusted self-propulsion speed of the tracer drops by a factor of $10$
at the highest densities studied, i.e., from about $\SI{1}{\micro\meter\per\second}$ to $\SI{0.1}{\micro\meter\per\second}$. This can be rationalized by the
expectation that due to deflections from the host-system particles,
the self-propulsion mechanism of the tracer in the experient becomes less
effective at fixed energy input with increasing density. This can refer
possibly to the laser-energy deposition onto the cap of the Janus particle
being less efficient, but also to the fact that the Janus particle requires
a region of reversible solvent-phase-separation induced by heating the cap
and close to it; the presence of other particles could well perturb this
fluid pattern to make self propulsion less strong.
Interestingly, the adjusted $v_0^s$ in the simulation display a maximum
at $\phi=0.76$. It is not evident where such a non-monotonic effectiveness
of the experimental driving would come from.

This peculiarity aside, both the experiment and the simulation data indicate
an interesting effect in the glass: although the host system is, over the
time scales that we can access, effectively arrested, the \gls{MSD}
of the active tracer continues to increase beyond the corresponding
cageing length scale. This could indicate a delocalization transition
of a strongly driven active tracer. Such delocalization is known for
passive but externally driven particles in a glass; a setup referred to
as active microrheology \cite{Puertas.2014}. It is not a priori evident whether
the same physical mechanisms applies in both cases: the externally driven
tracer is infinitely persistent in its motion if the external force
is kept constant, while the \gls{ABP} has only finite persistence time.
Even the limit $D_r^s\to0$ might not commute with the limit $t\to\infty$ taken
to decide whether the tracer becomes ultimately delocalized.

\subsection{Stokes-Einstein Relation}\label{sec:se}

\begin{figure}
\includegraphics[width=\linewidth]{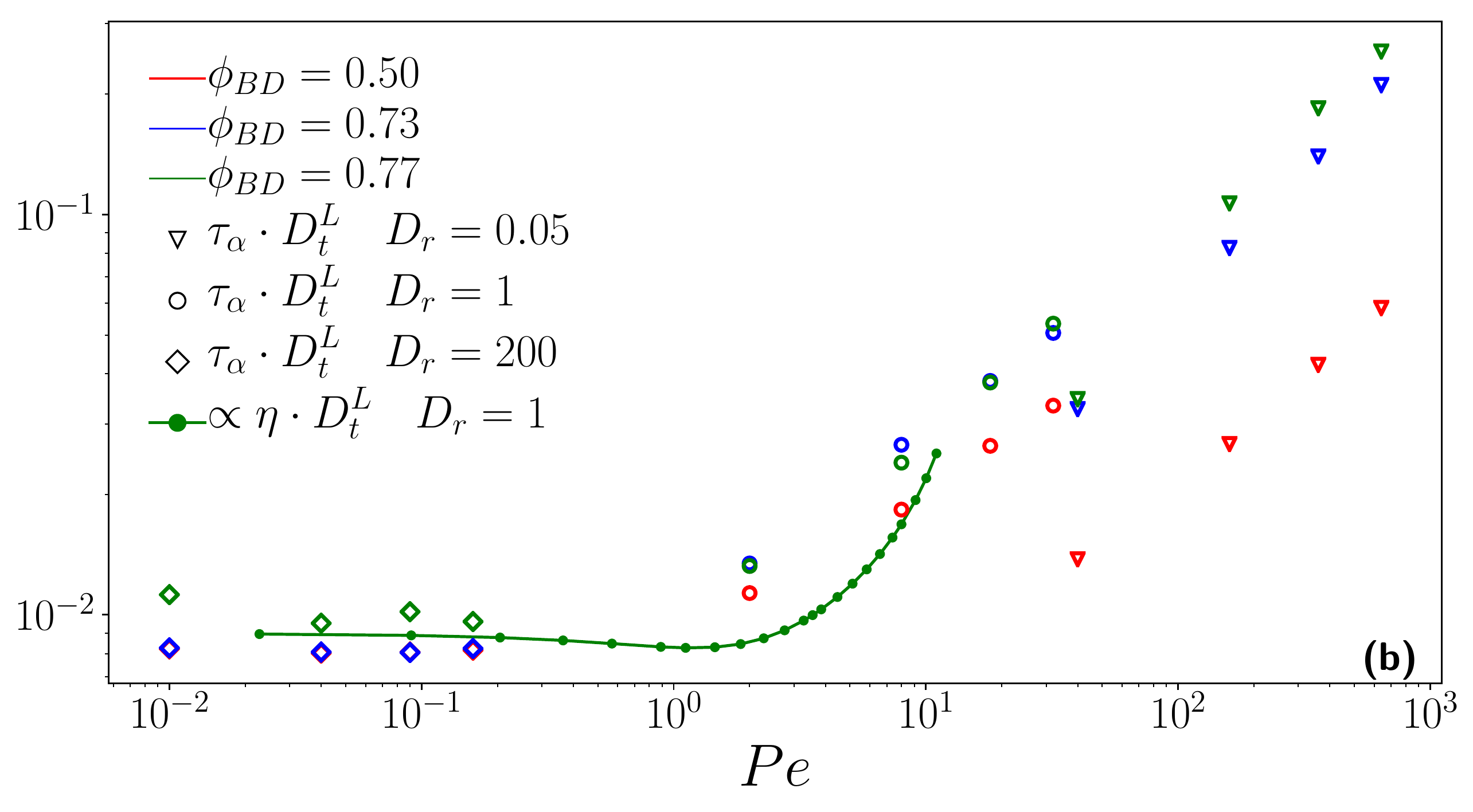}
\caption{\label{fig:447}
  Check of a generalized Stokes-Einstein relation between the active tracer
  long-time self-diffusion coefficient $D^L_t$ and the structural
  relaxation time $\tau_\alpha$ of the active host system. The product
  $D^L_t\tau_\alpha$ is shown for various simulations with different
  $D_r$ and at different packing fractions (symbols and colors as labeled),
  and for MCT with $D_r=1=D_r^s$ (line; for a state close to the glass
  transition) as a function of
  P\'eclet number $\Pe=v_0^2/2D_rD_t$.
}
\end{figure}

A prominent relation to link the tracer motion to dynamical features of
the host system that has been discussed in the context of the glass
transition is the generalized \gls{SE} relation.
Named after the famous result for the diffusion coefficient $D$ of a large
colloidal particle moving in a continuum fluid of viscosity $\eta$,
$D\eta\sim kT/\sigma$, the \gls{SE} relation in the context of glassy
dynamics refers to the fact that in the fluid regime described by
\gls{MCT}, both the tracer-diffusion coefficient and the host-fluid
viscosity are governed by the same cage relaxation processes, and hence
show the same control-parameter dependence asymptotically, even if the
tracer is of the same type as the particles comprising the host system
(and not infinitely larger as in the original \gls{SE} relation).

\Gls{MCT} explains the appearance of a \gls{SE} relation in this sense;
but it is also considered one of the theory's greatest failures to
not predict the violation of the \gls{SE} relation that is observed in
computer simulation for densities very close to or above the
\gls{MCT} transition point. Such violations are seen as indicative of
relaxation processes ``beyond'' \gls{MCT}.

The \gls{SE} relation is also the basis of the technique of passive
microrheology that aims to assess the rheological properties of the
host suspension by monitoring the displacement dynamics of an embedded
tracer particle. Microrheology has advantages over conventional rheology
when providing a sufficient amount of host-suspension fluid is not
feasible, and this makes it an interesting technique in particular in
the context of biofluids.

It is therefore interesting to check the applicability of the generalized
\gls{SE} relation in the active system. In the spirit of the theory,
we extract from our computer simulations the quantity $D^L_t\tau_\alpha$,
were $D^L_t$ is the long-time self-diffusion coefficient of an active
tracer in an active host suspension whose structural relaxation time
(as measured through the density correlation functions at finite $q$
close to the main peak of the static structure factor) is $\tau_\alpha$.
Figure~\ref{fig:447} shows the result for various densities, in simulations
with different persistence times $1/D_r^s=1/D_r$, as a function of
P\'eclet number. We observe that in the strongly active system,
the product $D\tau_\alpha$ increases by more than an order of magnitude,
and there is also a nontrivial $D_r$- and density-dependence.
Results from \gls{ABPMCT} are also shown (lines in Fig.~\ref{fig:447});
for these we have evaluated the \gls{ITT} expression for the linear-response
viscosity of the active host system \cite{julianphd} to calculate
$D^L_t\eta$. In the comparison with simulation, we have made use of the
fact that the Green-Kubo integral that determines the viscosity $\eta$ is,
close to the glass transition, dominated by a constant (the plateau value
of the dynamical correlation function in the cage regime) times the
structural relaxation time $\tau_\alpha$, so that a constant rescaling
can be performed to match the ``proper'' $D^L_t\eta$ to the more common
$D^L_t\tau_\alpha$ obtained in the simulation.
In the comparison we have also included a rescaling of the numerical value
of $v_0$ used in the theory, $v_0^\text{MCT}=1.5v_0^\text{BD}$;
this accounts for a known underestimation of the strength of the self-propulsion
forces in the \gls{MCT} vertex quantifying the collective relaxation of
density fluctuations and has been establishes in detail in comparison
to simulations of the density correlation functions
\cite{tagged}.
The \gls{ABPMCT} curve for $D_r=1$ then qualitatively agrees with the
simulation data; in particular the theory captures the increase of
$D^L_t\tau_\alpha$ with increasing $\Pe$.

This results emphasizes that the appearance of a generalized \gls{SE}
relation in \gls{MCT} is not trivial. While the fact that both $1/D$ and
$\eta$ asymptotically follow the same power law close to the glass transition
in the theory ensures that the product $D\eta$ approaches a constant; however,
the fact that this product results in an order of magnitude that is
comparable with simulation and moreover indicates a reaonsable ``effective
hydrodynamic radius'' of the tracer particle, is in a sense a numerical
coincidence.

Qualitatively, the increase with $\Pe$ that is seen in Fig.~\ref{fig:447}
can be rationalized: as the tracer particle becomes more active, it will
find it easier to diffuse and hence it will appear effecitvely ``smaller''
as long as the collective speeding up of the host-system relaxation is
less strong. Also, increasing the persistence length of the tracer particle
leads to more effective diffusion as measured through the generalized
\gls{SE} relation.

\subsection{Swim Velocities}\label{sec:vswim}

As a genuinely non-equilibrium transport coefficient, the effective
swim velocity $v(\phi)$ of \gls{ABP} plays a significant role. It
describes the density-renormalized propulsion, recognizing the fact
that the bare self-propulsion velocity $v_0$ of a free \gls{ABP}
is reduced due to interactions with the host-system particles.
We next asses the \gls{ITT} formula Eq.~\eqref{eq:vswimitt} with the
specific \gls{ABPMCT} closure, Eq.~\eqref{eq:vswimmct}, in comparison
with computer simulation.

\begin{figure}
\includegraphics[width=\linewidth]{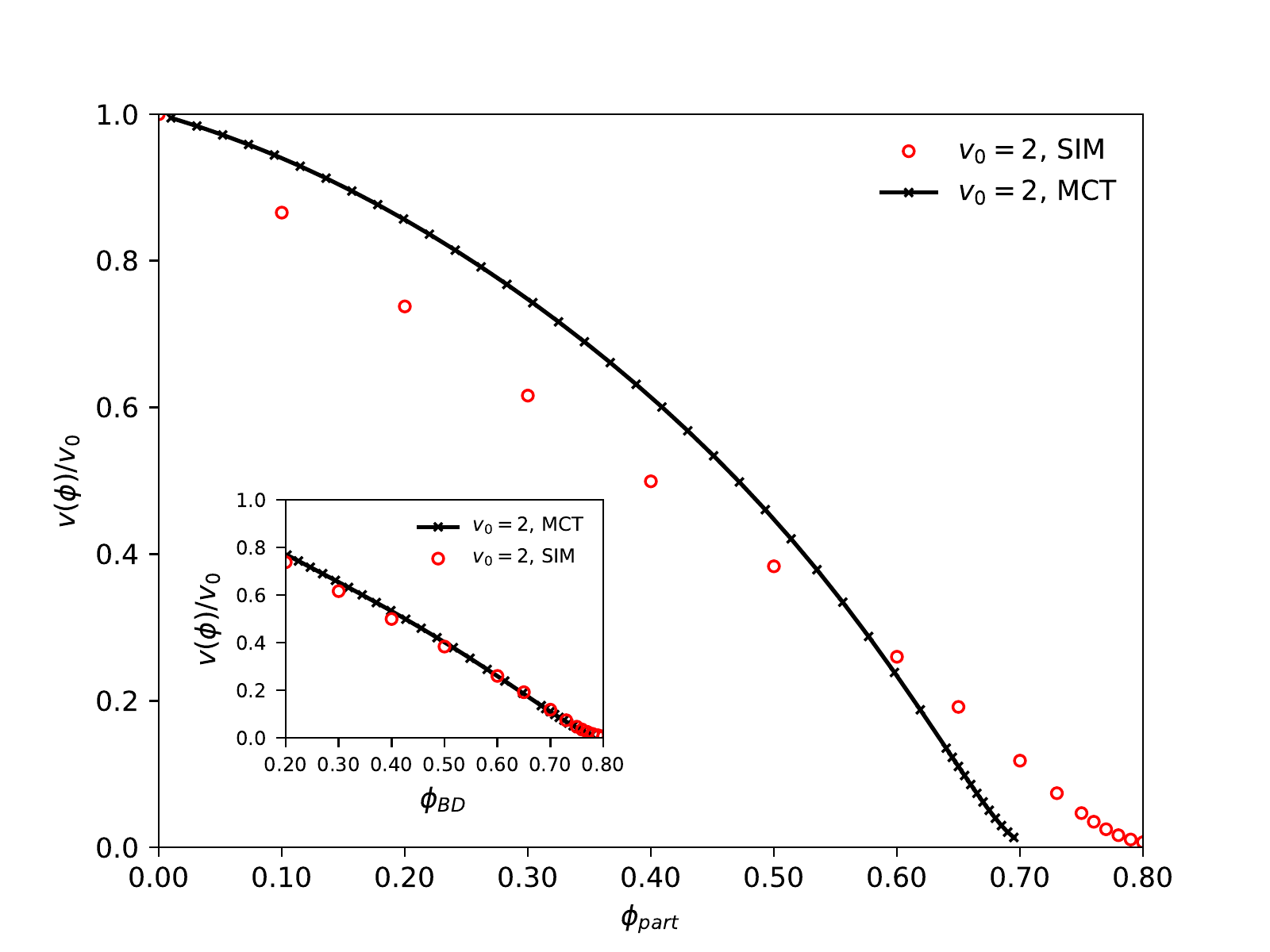}
\caption{\label{fig:457}
  Effective swim velocity $v(\phi)$ of interacting active Brownian particles,
  in units of the bare self-propulsion velocity $v_0=2$ of a single ABP,
  at packing fraction $\phi$. Symbols are from BD computer simulations.
  A line indicates parameter-free results from MCT.
  The inset shows the same data, but with the density in the MCT calculations
  adjusted to provide a good description of the relaxation time of the
  passive system.
}
\end{figure}

Figure~\ref{fig:457} shows the results from computer simulation and from
\gls{ABPMCT} for the density-dependent swim velocity in a fully active
system at a range of packing fractions spanning from the dilute system
to the glass transition. As expected, the swim velocity in the \gls{BD}
simulations is seen to decrease monotonically as interactions become more
important with increasing density. At the glass transition, long-range
motion of the particles ceases, and this implies that also the swim velocity
decays to zero.

As has been noted in similar simulations of soft-core systems before
\cite{Fily.2012,Sharma.2016}
the swim velocity almost follows a linear decrease with increasing density,
although closer inspection shows that in particular at high densities
there are some deviations from a linear law. The linear law is also
what has been assumed in continuum models to study \gls{MIPS}
\cite{Cates.2015}.

The \gls{ABPMCT} results at first sight are quite different. The theory
correctly predicts a monotonic decay with increasing density, and the
fact that the swim velocity approaches zero at the glass transition.
But the quantitative agreement with simulations is not optimal.
One should however note that in this direct comparison, no adjustment of
parameters in the theory has been made; in particular, as discussed above
in connection with the \gls{MSD}, one should expect that the theory
results need to be compared to simulations at a slightly different packing
fraction in order to account for the known numerical error of \gls{MCT}
in prediction the value of the glass-transition point $\phi_c$. In fact it is instructive to
readjust the packing-fraction axis such that the \emph{passive} \gls{MCT}
for each value of $\phi_\text{BD}$, quantitatively matches the relaxation
time of the density correlation functions. While close to $\phi_c$ this
results in a linear shift of packing fractions, outside the asymptotic
regime, quadratic terms in $\phi_\text{MCT}(\phi_\text{BD})$ are needed.

Letting aside the question of how to justify such mapping in detail, it
allows us to disentangle two very different effects: one of the quality
of the \gls{MCT} factorization in describing the cage effect, and
on of the quality of the \gls{ITT} application joint with \gls{ABPMCT}
in the formula for the swim velocity. Indeed, after adjusting the theory
to match the passive relaxation dynamics, also the swim velocities are
in rather good quantitative agreement with our simulations, as demonstrated
in the inset of Fig.~\ref{fig:457}.

\begin{figure}
\includegraphics[width=\linewidth]{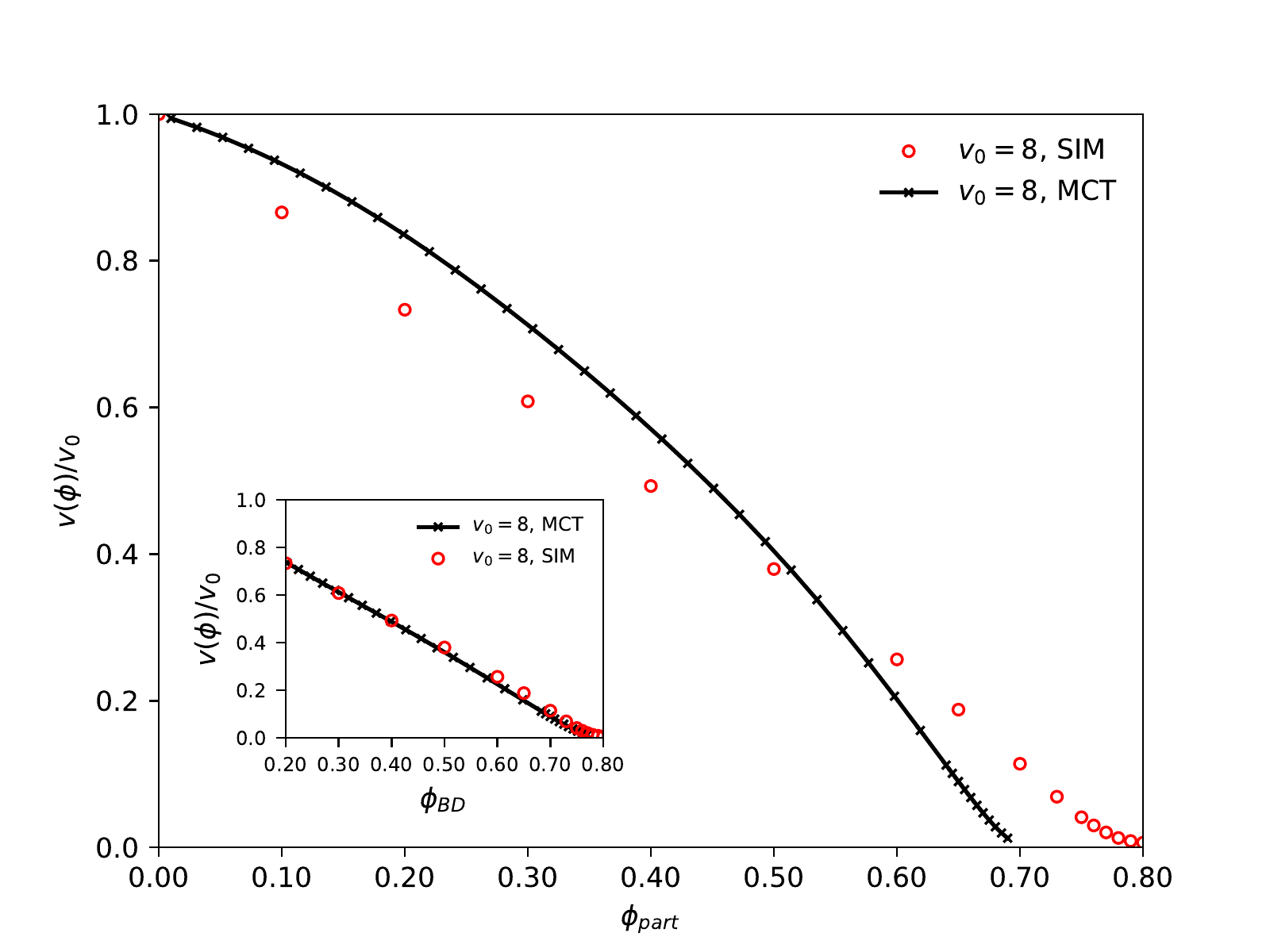}
\caption{\label{fig:439}
  Effective swim velocity $v$ of a single ABP tracer in a passive host suspension
  of packing fraction $\phi$, in units of the tracer's bare
  self-propulsion velocity $v_0^s=8$. Symbols are computer simulation results,
  lines are from MCT. The inset shows the same comparison with a density
  axis that is adjusted from fitting MCT to the passive host system's
  relaxation dynamics.
}
\end{figure}

A qualitatively similar finding also holds for the case of a single
\gls{ABP} tracer particle that is embedded in a passive host suspension,
Fig.~\ref{fig:439}. Here, the theory predicts -- after the mapping of
densities described above -- the simulation results quantiatively up to
swimming speeds of around $v_0^s=8$; higher values are currently out of reach
for the numerical algorithm solving the \gls{MCT} equations of motion.
For the active bath, a velocity of $v_0=8$ results in a qualitatively
similar curve for $v(\phi)$, but here, somewhat stronger deviations are
seen between theory and simulations at medium densities.

The striking observation is that without parameter adjustment,
\gls{ABPMCT} predicts the swim velocity to decay initially quadratically with
increasing density, not linearly. This is a feature of the way the \gls{MCT}
approximation is constructed: the fluctuating interaction forces are
expressed through terms that are quadratic in the fluctuating densities.
At low $\phi$, this is not adequate.
%an this has already been remarked in
%passive suspensions of colloids mixed with free polymer, where the low-density
%regime of kinetic arrest becomes interesting \cite{bergenholtz}.
As our mapping in the insets of Figs.~\ref{fig:457} and \ref{fig:439}
shows, the calculation of the swim velocity inherits this deficiency of
\gls{MCT}, i.e., it is mainly the incorrect density-dependence of the
low-density relaxation dynamics of density fluctuations that also leads
to prima facie incorrect descriptions of the swim velocities.

\section{Conclusion}\label{sec:conclusion}

We have reviewed the development of the mode-coupling theory for active
Brownian particles to describe the mean-squared displacements of an active
tracer in a glass-forming host system. The theory was shown to compare
favorably to computer-simulation results for the hard-disk system,
at moderate P\'eclet numbers (up to $\Pe\approx640$). It describes
the sequence of sub- and super-diffusive motion observed in the \gls{MSD}
of \gls{ABP} embedded in a dense host system, and rationalizes it as
arising from a competition of the time scales of nearest-neighbor
cageing with those of persistent motion.

A direct comparison between \gls{ABPMCT} and current experimental data
on the self-propelled tracer motion of Janus particles in a glass-forming
colloidal suspension was unfortunately not possible. Numerical instabilities
in solving the \gls{ABPMCT} equations of motion prevent us from addressing
the regime of extremely strong self-propulsion. (From the experimental data
one estimates $\Pe^s\gtrsim10^3$ at intermediate densities.)
It remains to be seen whether
improved numerical schemes and/or computational efforts to increase
the agnular-mode cutoff employed in the numerics will remedy this situation.

However, we have been able to directly compare \gls{EDBD} simulations for
active-tracer motion in a host suspension of passive hard disks with
experiment. It appears that in experiment, both the rotational diffusion
and the self-propulsion velocity of the active tracer depend sensitively
on the vicinity to the host-suspension glass transition. These effects
are not included in the common \gls{ABP} model of self-propelled particles,
and complicate the analysis. The strong change in rotational diffusion
was noted in experiment directly \cite{Lozano.2019}, and attributed to
viscoelastic coupling with the host system. If true, this would necessitate
an approach where the orientational motion of the \gls{ABP} tracer
couples to the collective density fluctuations of the host system;
the development of a fully microscopic theory for this situation
needs to be left for future work.

For the regime that is accessible within the theory, non-equilibrium
transport coefficients
such as the long-time tracer-diffusion coefficient and the effective
swim velocity are predicted.
One result is the deviation from the commonly assumed Stokes-Einstein
relation with increasing $\Pe$; it shows that in extracting quantitative
information on the host-system viscosity from the long-time diffusivity
of a tracer in active fluids, one needs to be careful.

Our comparison with simulation for the swim velocities demonstrates
that a decisive factor in predicting these correctly is the correct
modeling of the structural relaxation time of the density fluctuation
dynamics already in the passive host system. \Gls{MCT} is in quantitative
error here, and in particular a low densities it does not yield the
correct leading-order variation with packing fraction. This error is often
overlooked, because usual comparisons of \gls{MCT} with simulation or
experiment focus on the vicinity of the glass transition, for which the theory
was designed. It becomes apparent when calculating for example the
density-dependent swim velocity of \gls{ABP} with the theory: while at
high densities, the approach of $v(\phi)$ to zero as $\phi\to\phi_c$
is correctly captured, the overall shape of the $v(\phi)$-vs-$\phi$ curve
is quite different from the almost linear variation that one finds in
simulation. We have demonstrated here that this difference does not indicate
a failure of the \gls{ABP}-extension of \gls{MCT} per se, but rather
that \gls{ABPMCT} inherits a deficiency from the original \gls{MCT} that
one needs to account for.

The results shown in Figs.~\ref{fig:457} and \ref{fig:439} nevertheless confirm
a peculiar approach inherent to \gls{MCT} to such generalized Green-Kubo
relations: while the original expression derived using \gls{ITT},
Eq.~\eqref{eq:vswimittorig} describes the reduction of the swim velocity
due to interactions by a subtractive mobility term, at high densities an approach
that translates the Green-Kubo epxression into an additive friction term,
Eq.~\eqref{eq:vswimitt}. In particular it allows to explain that the
effective swim velocity appears to vanish at the glass transition, as also
observed in our \gls{EDBD} simulations.

Keeping this in mind, our recent extension of \gls{ABPMCT} to include
various types of tracer particles is readily generalized
to binary mixtures of active and passive particles
\cite{mixtures}. This should then provide a promising microscopic theory to
address the question how the addition of a few active particles to a passive
suspension speeds up the dynamics, or how interacting passive tracers
experience enhanced diffusion due to being embedded in an active fluid.

\subsection*{Acknowledgments}

This project was funded by Deutsche Forschungsgemeinschaft (DFG),
through project Vo~1270/7-2 in the context of the
Special Priority Programme SPP~1726 ``Microwsimmers''.
We acknowledge fruitful discussions with C.~Bechinger, C.~Lozano,
H.~L\"owen.
We also thank C.~Lozano for providing the experimental data, and
S.~Mandal for invaluable help with the computer-simulation code.

\subsection*{Authors contributions}
JR and ThV performed the theory calculations; JR implemented the
numerics and performed the simulations to obtain the
mean-squared displacements, and performed the fit to experimental data.
LG implemented the calculation of swim velocities in the simulation and
performed the comparison to theory.
ThV devised the research project and wrote the paper.
All the authors have read, discussed, and approved the final manuscript.

\bibliographystyle{epj}
\bibliography{lit}

\end{document}